\documentstyle[11pt,aaspp4]{article}

\slugcomment{Accepted by The Astronomical Journal}

\lefthead{Schneider et al.}
\righthead{SDSS Quasar Catalog I.}

\begin{document}

\title{The Sloan Digital Sky Survey Quasar Catalog~I. Early Data Release
\footnote{Based on observations obtained with the Sloan Digital Sky
Survey, which is owned and operated by the Astrophysical Research Consortium.}
}

\author{
Donald~P.~Schneider\altaffilmark{\ref{PennState}},
Gordon~T.~Richards\altaffilmark{\ref{PennState}},
Xiaohui~Fan\altaffilmark{\ref{IAS}},
Patrick~B.~Hall\altaffilmark{\ref{Princeton}}$^,$\altaffilmark{\ref{CChile}},
Michael~A.~Strauss\altaffilmark{\ref{Princeton}},
Daniel~E.~Vanden~Berk\altaffilmark{\ref{FNAL}},
James~E.~Gunn\altaffilmark{\ref{Princeton}},
Heidi~Jo~Newberg\altaffilmark{\ref{RPI}},
Timothy~A.~Reichard\altaffilmark{\ref{PennState}},
C.~Stoughton\altaffilmark{\ref{FNAL}},
Wolfgang~Voges\altaffilmark{\ref{MPE}},
Brian~Yanny\altaffilmark{\ref{FNAL}},
Scott~F.~Anderson\altaffilmark{\ref{UW}},
James~Annis\altaffilmark{\ref{FNAL}},
Neta~A.~Bahcall\altaffilmark{\ref{Princeton}},
Amanda~Bauer\altaffilmark{\ref{Cincinnati}},
Mariangela~Bernardi\altaffilmark{\ref{Chicago}},
Michael~R.~Blanton\altaffilmark{\ref{FNAL}},
William~N.~Boroski\altaffilmark{\ref{FNAL}},
J.~Brinkmann\altaffilmark{\ref{APO}},
John~W.~Briggs\altaffilmark{\ref{Chicago}},
Robert~Brunner\altaffilmark{\ref{Caltech}},
Scott~Burles\altaffilmark{\ref{Chicago}},
Larry~Carey\altaffilmark{\ref{UW}},
Francisco~J.~Castander\altaffilmark{\ref{Chile}},
A.J.~Connolly\altaffilmark{\ref{Pittsburgh}},
Istv\'an~Csabai\altaffilmark{\ref{JHU}}$^,$\altaffilmark{\ref{Hungary}},
Mamoru~Doi\altaffilmark{\ref{Tokyo2}},
Scott~Friedman\altaffilmark{\ref{JHU}},
Joshua~A.~Frieman\altaffilmark{\ref{Chicago}},
Masataka~Fukugita\altaffilmark{\ref{CosJapan}}$^,$\altaffilmark{\ref{IAS}},
Timothy~M.~Heckman\altaffilmark{\ref{JHU}},
G.S.~Hennessy\altaffilmark{\ref{USNODC}},
Robert~B.~Hindsley\altaffilmark{\ref{NRL}},
David~W.~Hogg\altaffilmark{\ref{IAS}},
\v Zeljko~Ivezi\'c\altaffilmark{\ref{Princeton}},
Stephen~Kent\altaffilmark{\ref{FNAL}},
Gillian~R.~Knapp\altaffilmark{\ref{Princeton}},
Peter~Z.~Kunszt\altaffilmark{\ref{JHU}},
Donald~Q.~Lamb\altaffilmark{\ref{Chicago}}$^,$\altaffilmark{\ref{Fermi}},
R.~French~Leger\altaffilmark{\ref{UW}},
Daniel~C.~Long\altaffilmark{\ref{APO}},
Jon~Loveday\altaffilmark{\ref{Sussex}},
Robert~H.~Lupton\altaffilmark{\ref{Princeton}},
Bruce~Margon\altaffilmark{\ref{STScI}},
Avery~Meiksin\altaffilmark{\ref{Edinburgh}},
Aronne~Merelli\altaffilmark{\ref{Caltech}},
Jeffrey~A.~Munn\altaffilmark{\ref{USNOAZ}},
Matthew~Newcomb\altaffilmark{\ref{CMU}},
R.C.~Nichol\altaffilmark{\ref{CMU}},
Russell~Owen\altaffilmark{\ref{UW}},
Jeffrey~R.~Pier\altaffilmark{\ref{USNOAZ}},
Adrian~Pope\altaffilmark{\ref{CMU}},
Constance~M.~Rockosi\altaffilmark{\ref{Chicago}},
David~H.~Saxe\altaffilmark{\ref{IAS}},
David~Schlegel\altaffilmark{\ref{Princeton}},
Walter~A.~Siegmund\altaffilmark{\ref{UW}},
Stephen~Smee\altaffilmark{\ref{JHU}},
Yehuda~Snir\altaffilmark{\ref{CMU}},
Mark~SubbaRao\altaffilmark{\ref{Chicago}},
Alexander~S.~Szalay\altaffilmark{\ref{JHU}},
Aniruddha~R.~Thakar\altaffilmark{\ref{JHU}},
Alan~Uomoto\altaffilmark{\ref{JHU}},
Patrick~Waddell\altaffilmark{\ref{UW}},
and
Donald~G.~York\altaffilmark{\ref{Chicago}}$^,$\altaffilmark{\ref{Fermi}}
}


\newcounter{address}
\setcounter{address}{2}
\altaffiltext{\theaddress}{Department of Astronomy and Astrophysics, The
   Pennsylvania State University, University Park, PA 16802.
\label{PennState}}
\addtocounter{address}{1}
\altaffiltext{\theaddress}{The Institute for Advanced Study, Princeton,
   NJ 08540.
\label{IAS}}
\addtocounter{address}{1}
\altaffiltext{\theaddress}{Princeton University Observatory, Princeton,
   NJ 08544.
\label{Princeton}}
\addtocounter{address}{1}
\altaffiltext{\theaddress}{Pontificia Universidad Cat\'olica de Chile,
   Departamento de Astronom\'{\i}a y Astrof\'{\i}sica, Facultad de F\'{\i}sica,
   Casilla~306, Santiago~22, Chile.
\label{CChile}}
\addtocounter{address}{1}
\altaffiltext{\theaddress}{Fermi National Accelerator Laboratory, P.O. Box 500,
   Batavia, IL 60510.
\label{FNAL}}
\addtocounter{address}{1}
\altaffiltext{\theaddress}{Department of Physics, Applied Physics and
   Astronomy, Rensselaer Polytechnic Institute, Troy, NY 12180.
\label{RPI}}
\addtocounter{address}{1}
\altaffiltext{\theaddress}{Max-Planck-Institute f\"ur extraterrestrische
   Physik, Geissenbachstr.~1, D-85741 Garching, Germany.
\label{MPE}}
\addtocounter{address}{1}
\altaffiltext{\theaddress}{University of Washington, Department of
   Astronomy, Box 351580, Seattle, WA 98195.
\label{UW}}
\addtocounter{address}{1}
\altaffiltext{\theaddress}{Department of Physics, University of Cincinnati,
   400~Physics~Bldg., Cincinati, OH~45221.
\label{Cincinnati}}
\addtocounter{address}{1}
\altaffiltext{\theaddress}{Astronomy and Astrophysics Center, University of
   Chicago, 5640 South Ellis Avenue, Chicago, IL 60637.
\label{Chicago}}
\addtocounter{address}{1}
\altaffiltext{\theaddress}{Apache Point Observatory, P.O. Box 59,
   Sunspot, NM 88349-0059.
\label{APO}}
\addtocounter{address}{1}
\altaffiltext{\theaddress}{Astronomy Department, California Institute of
   Technology, Pasadena, CA 91125.
\label{Caltech}}
\addtocounter{address}{1}
\altaffiltext{\theaddress}{Universidad de Chile, Casilla 36-D, Santiago,
   Chile.
\label{Chile}}
\addtocounter{address}{1}
\altaffiltext{\theaddress}{Department of Physics and Astronomy, University
   of Pittsburgh, 3941 O'Hara Street, Pittsburgh, PA 15260.
\label{Pittsburgh}}
\addtocounter{address}{1}
\altaffiltext{\theaddress}{Department of Physics and Astronomy,
   The Johns Hopkins University, 3701 University Drive, Baltimore, MD 21218.
\label{JHU}}
\addtocounter{address}{1}
\altaffiltext{\theaddress}{Department of Physics of Complex Systems,
   E\"otv\"os University, P\'azm\'ay P\'eter \hbox{s\'et\'any 1/A,}
   H-1117, Budapest, Hungary.
\label{Hungary}}
\addtocounter{address}{1}
\altaffiltext{\theaddress}{Department of Astronomy and Research Center for the
   Early Universe, School of Science, University of Tokyo, Mitaka,
   Tokyo 181-0015, Japan.
\label{Tokyo2}}
\addtocounter{address}{1}
\altaffiltext{\theaddress}{Institute for Cosmic Ray Research, University
   of Tokyo, Kashiwa, 2778582, Japan.
\label{CosJapan}}
\addtocounter{address}{1}
\altaffiltext{\theaddress}{US Naval Observatory, 3450 Massachusetts Avenue NW,
   Washington, DC 20392-5420.
\label{USNODC}}
\addtocounter{address}{1}
\altaffiltext{\theaddress}{Remote Sensing Division, Code~7215, Naval Research
   Laboratory, 4555~Overlook~Ave. SW, Washington, DC~20375.
\label{NRL}}
\addtocounter{address}{1}
\altaffiltext{\theaddress}{The University of Chicago, Enrico Fermi Institute,
  5640 South Ellis Avenue, Chicago, IL 60637.
\label{Fermi}}
\addtocounter{address}{1}
\altaffiltext{\theaddress}{Astronomy Centre, University of Sussex, Falmer,
   Brighton BN1~9QJ, UK.
\label{Sussex}}
\addtocounter{address}{1}
\altaffiltext{\theaddress}{Space Telescope Science Institute,
   3700 San Martin Drive, Baltimore, MD 21218.
\label{STScI}}
\addtocounter{address}{1}
\altaffiltext{\theaddress}{
Royal Observatory, Edinburgh, Blackford Hill,
   Edinburgh EH9~3HJ, UK.
\label{Edinburgh}}
\addtocounter{address}{1}
\altaffiltext{\theaddress}{US Naval Observatory, Flagstaff Station,
   P.O. Box 1149, Flagstaff, AZ 86002-1149.
\label{USNOAZ}}
\addtocounter{address}{1}
\altaffiltext{\theaddress}{Dept. of Physics, Carnegie Mellon University,
     5000~Forbes Ave., Pittsburgh, PA~15232.
\label{CMU}}


\vbox{
\begin{abstract}
We present the first edition of the Sloan Digital Sky Survey (SDSS)
Quasar Catalog.  The catalog consists of the~3814 objects (3000 discovered
by the SDSS) in the
initial SDSS public data release that have at least one emission line with
a full width at half maximum larger than~1000~km~s$^{-1}$, luminosities
brighter than \hbox{$M_{i^*} = -23$,}  and highly reliable redshifts.
The area covered by the catalog is 494~deg$^2$; the
majority of the objects were found
in SDSS commissioning data using a multicolor selection technique. 
The quasar redshifts range from~0.15 to~5.03.
For each object the catalog presents positions accurate
to better than~0.2$''$~rms per coordinate,
five band ($ugriz$) CCD-based photometry with typical accuracy
of~0.05~mag, radio and X-ray emission properties, and
information on the morphology and selection method.  Calibrated spectra
of all objects in the catalog, covering the wavelength region~3800
to~9200~\AA\ at a spectral resolution \hbox{of 1800-2100,} are also available.
Since the quasars were selected during the commissioning period, a time
when the quasar selection algorithm was undergoing frequent revisions, the
sample is not homogeneous and is not intended for statistical analysis.

\end{abstract}
}

\keywords{catalogs, surveys, quasars:general}

\section{Introduction}

Since the first measurement of a quasar redshift nearly 40~years
ago (Schmidt~1963), the number of known quasars has steadily risen;
the NASA/IPAC Extragalactic Database (NED) and the quasar catalog of
V\'eron \& Veron~(2001) each contain
on the order of~25,000 quasars.  A large fraction of these objects were
recently discovered by the 2dF Quasar Survey (Croom et al.~2001), which
is extremely effective at identifying quasars with redshifts below~3.0
that have $b_J$ magnitudes between~18.25 and~20.85.

This paper presents the first edition of the Sloan Digital Sky Survey
(SDSS) Quasar Catalog.  The goal of the SDSS quasar survey is to
obtain spectra of~$\approx$~100,000 quasars from 10,000~deg$^2$ of the
North Galactic Cap.  For quasars with \hbox{$(u-g) < +1.5$} (which corresponds
to a redshift of near three),
the survey will reach a flux limit of~$i$~$\approx$~19;
quasars with \hbox{$(u-g) > +1.5$} (corresponding to a redshift range
in the SDSS of between 3.0 and~$\approx$~5.5-6.0), the sensitivity limit
will be~$i$~$\approx$~20.  The survey will provide CCD-based photometry
in five broad bands covering the entire optical window,
morphological information, and spectra from
\hbox{3800 \AA\ to 9200 \AA\ at} a spectral resolution \hbox{of 1800-2100.}
A review of the SDSS is given by York et al.~(2000); Richards et al.~(2002)
present the details of the quasar target algorithm.

The catalog in the present paper
consists of the 3814 objects in the SDSS Early Data Release
(EDR; Stoughton et al.~2002) with reliable redshifts whose spectra
have at least one emission line with a FWHM
broader \hbox{than 1000 km s$^{-1}$} and which have a luminosity larger than
\hbox{$M_{i^*} = -23$}  (calculated assuming an
\hbox{$H_0$ = 50 km s$^{-1}$ Mpc$^{-1}$,} 
\hbox{$\Omega_M$ = 1.0,} 
\hbox{$\Omega_{\Lambda}$ = 0} cosmology, which will be used throughout
this paper).  The quasars range in redshift from~0.15 to~5.03, and
3000~(79\%) were discovered by the SDSS (an object is classified as
previously known if NED contains a quasar within~5$''$ of the SDSS position).

A few quasar-related
studies based on subsets of the SDSS Early Release
Data have been recently published.
The most comprehensive investigations, both using
samples containing more than~2000 quasars, are the
quasar color-redshift relation for redshifts between zero and
five (Richards et al.~2001a) and the construction of
a very high signal-to-noise ratio composite quasar spectrum
(Vanden~Berk et al.~2001).
The SDSS has proven to be extremely effective at identifying high-redshift
quasars; to date the SDSS has discovered more than~140 at redshifts
above four, and ten of the eleven known quasars at \hbox{$z \ge 5$} 
(see Zheng et al.~2000, Anderson et al.~2001, Fan et al.~2001,
and references therein; the sole non-SDSS $z>5$ quasar is described in
Sharp et al.~2001).

The observations used to produce the catalog are presented in
Section~2.  The construction of the catalog and the catalog format
are discussed in Sections~3 and~4, respectively, and Section~5
contains a summary of the catalog.  A brief discussion of plans for
future editions of the catalog is given in Section~6.  The catalog material
can be found at a public web site
\footnote{\tt http://archive.stsci.edu/sdss/documents/prepared\_ds.html}.

\section{Observations}

\subsection{Sloan Digital Sky Survey}

The Sloan Digital Sky Survey
uses a CCD camera \hbox{(Gunn et al. 1998)} on a
dedicated 2.5-m telescope
at Apache Point Observatory,
New Mexico, to obtain images in five broad optical bands over
10,000~deg$^2$ of the high Galactic latitude sky centered approximately
on the North Galactic Pole.  The five filters 
(designated $u$, $g$, $r$, $i$, and~$z$) 
cover the entire wavelength range of the CCD
response \hbox{(Fukugita et al.~1996);} the filter response
curves are given in Stoughton et al.~(2002).
Since the SDSS photometric system is not yet
finalized, we refer to the SDSS photometry presented here as
$u^*$, $g^*$, $r^*$, $i^*$, and~$z^*$.  The photometric calibration is
reproducible to
0.05, 0.03, 0.03, 0.03, and~0.05
magnitudes in $u^*$, $g^*$, $r^*$, $i^*$, and~$z^*$, respectively; the
absolute calibration in Janskys is uncertain at the~10\% level.
All magnitudes in the quasar catalog refer to the point spread function
measurements of the photometric pipeline (see Stoughton et al.~2002 for
details).

Photometric calibration is provided by simultaneous
observations with a 20-inch telescope at the same site (see Hogg
et al.~2001 and Stoughton et al.~2002). The
survey data processing software measures the properties of each detected object
in the imaging data in all five bands, and determines and applies both
astrometric and photometric
calibrations (Pier et al., unpublished; \hbox{Lupton et al. 2001)}.
The image quality in the EDR, which consists of observations
taken during
the SDSS commissioning period, is considerably poorer than
that expected for the survey proper; the~95\% completeness limits for stars
in the EDR are
typically~22.0, 22.2, 22.2, 21.3,
and~20.5 in~$u^*$, $g^*$, $r^*$, $i^*$ and~$z^*$,
respectively.  The image of an unresolved source brighter
than \hbox{$r^* \approx 14$} will be saturated.

The imaging data in the SDSS EDR consists of eight imaging scans
(SDSS scan numbers~94, 125, 752, 756, 1336, 1339, 1356, and~1359),\
acquired between
September~1998 and April~2000, that cover approximately~500~deg$^2$.
The first four scans are along the celestial equator in the
Northern and Southern high-latitude Galactic sky; the final four
contain 68~deg$^2$ in the SIRTF First Look Survey region 
(see Stoughton et al.~2002).

\subsection{Target Selection}

The SDSS filter system was designed to allow quasars at redshifts between
zero and
six to be identified with multicolor selection techniques.  The effective
wavelength
of the~$u$ filter is shortward of the Balmer discontinuity;
therefore the color difference between low-redshift quasars
and early-type stars is larger in $(u-g)$ than the standard
color of~$(U-B)$.  The
inclusion of the near-infrared filter ($z$) extends the maximum redshift
for SDSS quasars
out to~$\approx$~6 (Fan et al.~2001).  The vast majority of quasars
follow a tight color-redshift relation in SDSS filters
(Richards et al.~2001a);
this feature allows development of techniques that may produce reliable
photometric redshifts for quasars (Richards et al. 2001b, Budavari et al.~2001).
In addition to the multicolor selection, unresolved objects brighter
\hbox{than $i \approx 19$} that are coincident with FIRST radio sources
(Becker, White, \&~Helfand~1995) are also identified as quasar
candidates.

Note that the point spread function magnitudes are used for the quasar
target selection.  For candidates whose likely redshifts are less than three,
both extended and point sources are included as quasar candidates; however,
extended sources are excluded if they lie in a region of color space that is
densely occupied by normal galaxies (see Richards et al.~2002).  At larger
redshifts, an object must be unresolved in the SDSS images
to become a spectroscopic target.

Target selection also imposes a maximum brightness limit on the objects.
Accurate photometry of point sources brighter than \hbox{$r \approx 14$}
is impossible as their images are saturated; objects that have saturated
pixels are dropped from further consideration.
An additional constraint is introduced to prevent
saturation and fiber cross-talk problems in the SDSS spectroscopic
observations; an object
cannot be included in the quasar spectroscopic program if
it has an~$i$  magnitude
brighter than~15.0.  Objects may also dropped
from consideration if the photometric measurements are considered
suspect, for example objects close to very bright stars, data affected
by cosmic rays, etc.

One of the most important tasks during the SDSS commissioning period
was to refine the quasar target selection algorithm.  This selection
algorithm was varied throughout the time that the EDR
observations were obtained, so the objects in this catalog were not
found via a uniform set of selection criteria.  Indeed, some of the
quasars in the catalog were not spectroscopically targeted just by
the SDSS quasar selection algorithm, but rather by one of the other
modules (Stoughton et al.~2002): galaxies, stars (mostly aimed at
objects with the colors of unusual types of stars), optical
counterparts of ROSAT sources, or serendipitous targets (again, 
objects of extreme colors, but defined independently from the quasar
selection color limits).  For a detailed discussion of the process
of spectroscopic target selection, see Stoughton et al.~(2002); Richards
et al.~(2002) discuss the final SDSS target selection criteria
for quasars. 

The evaluations of the survey selection efficiency (number of
quasars compared to
the number of quasar candidates) and completeness (fraction of quasars
found by the SDSS) are complex tasks, particularly when dealing with the
EDR data base; these issues, as concern future data releases,
will be addressed by Richards et al.~(2002).
The current estimate on the efficiency of the final selection algorithm
is 65-70\%; the algorithm's completeness, determined from simulations and
comparison with previously known quasars, should be approximately~90\%.
Both of these values have brightness and redshift dependences.

\subsection{Spectroscopy}

Spectroscopic targets chosen by the various SDSS selection algorithms
($e.g.,$ quasars, galaxies, stars, serendipity) are organized onto
a series of 3$^{\circ}$ diameter circular fields (Blanton et al.~2001).
The positions are mapped and drilled into aluminum plates; each plate
contains 640 fibers that feed two double
spectrographs mounted at the Cassegrain focus of the SDSS 2.5-m telescope
(see York et al.~2000 and Castander et al.~2001 for details).
The spectrographs produce data covering \hbox{3800--9200 \AA }, with the
beam split at~6150~\AA\ by a dichroic.  The data have a 
spectral resolution ranging from 1800 to 2100.
A total of 320 fibers enter each spectrograph;
each fiber subtends a diameter of~3$''$ on the sky, and because of mechanical
constraints the centers of the fibers must be separated by at
least~55$''$ (although in regions of sky in which plates overlap, one
can have spectra of objects separated by less than this angle).
Typically about~75 quasar
candidate spectra among the~640 fibers are observed in a
45-minute observation (broken into three 15-minute exposures) of a field; the
exposure time is increased in conditions of poor seeing or reduced sky
transparency to meet survey's minimum signal-to-noise ratio requirement
of~(S/N)$^2$ of~15 per spectrograph pixel \hbox{at $g^* = 20.2$}
\hbox{and $i^* = 19.9$.}  (See Stoughton et al.~2002 for an extensive
discussion of the spectroscopic observations.)

Observations from~92 spectroscopic fields are used to form the catalog.
The EDR contains~95 spectroscopic plates; of the three ``missing" plates,
one was a special observation of a star cluster, and the other two were
duplicates of other EDR plates but drilled for observations at different
airmasses.  The celestial locations of the~92 plates are
are displayed in
Figure~1.  The total area covered by the spectroscopic observations
is 494~deg$^2$ (as can be seen from the figure, there is significant
overlap between many of the fields).  The locations of the plate centers
are given in~Table~1, along with the number of quasars in the catalog
contained on each plate.  Note the wide range (10 to~123) in the number
of quasars per spectroscopic plate in these commissioning
observations; this variation is due to the various tests carried out during the
commissioning exercise.

The data, along with the associated calibration frames, are processed by
the SDSS Spectroscopic Pipeline (Burles et al., unpublished), which removes
instrumental effects, extracts the spectra, determines the wavelength
calibration, subtracts the sky spectrum, removes the atmospheric
absorption bands, and performs the flux calibration.  

The calibrated spectra are classified into various groups
(e.g., star, galaxy, quasar) by another automated software pipeline
(Frieman et al., unpublished).
The quasar classification is based
solely on the presence of broad emission lines in the spectra;
the classification software does not
employ information about the selection of the object
($e.g.$, was the spectrum obtained because the target selection process
identified the object as a quasar candidate?), nor is luminosity used by the
SDSS pipeline
as a criterion for designating an object as a quasar.
See Stoughton et al.~(2002) for details regarding the spectral classification
criteria.

The redshifts are measured by 
a combination of cross-correlation (using Fourier
techniques) to a
quasar template, and searches for emission-lines, together with code
that recognizes the onset of the Lyman $\alpha$ forest.
The software returns a redshift quality flag; when this flag indicates that
a reliable redshift cannot be assigned, the redshift is measured manually.
In practice, all of the quasar spectra presented herein (and the sample from
which they were drawn) were visually inspected multiple times.

Figure~2 shows the calibrated SDSS spectra of six of the catalog quasars
representing a wide range of properties;
all were previously unknown.  We discuss these individual objects in
Section~5.  These spectra have been slightly smoothed for display purposes.

\section{Construction of the SDSS Quasar Catalog}

The quasar catalog was constructed in three stages.  The first step,
which produced over~99\% of the entries in the catalog, was simply to
find the objects (4487 in total) 
in the EDR that the spectroscopic pipeline classified as quasars.
These objects were selected using a simple SQL query to
the EDR database using the SDSS Query
Tool\footnote{\tt http://archive.stsci.edu/sdss/software/} (Stoughton et
al.~2002).
We requested all
objects with spectral classifications of SPEC\_QSO
({\tt specClass=3}) or SPEC\_HIZ\_QSO ({\tt specClass=4}) from the
sxPrimary class of objects.  For example, in SQL notation:
\begin{verbatim}
SELECT objID
FROM sxPrimary
WHERE specobj.specClass == 3 || specobj.specClass == 4
\end{verbatim}

This query does not necessarily return {\em all} quasars in the EDR
database, but rather it identifies all spectra that meet some
well-defined criteria that cause them to belong to the sxPrimary class
of objects and to be classified as quasars.

This data base was supplemented by two additional efforts.  A total of
16 quasars, missed by the SDSS pipeline but identified during a visual
inspection of all the EDR spectra during a search for
extreme BAL quasars, were added to the quasar list.
Besides BAL quasars, these objects include
two star-quasar superpositions \hbox{(SDSS J012412.47$-$010049.8} and
\hbox{SDSS J014349.14+002128.4)} and the enigmatic object
\hbox{SDSS J010540.75$-$003314.0;} see Hall et al.~(2002) for a
discussion of this sample.

At a late stage of the production of the catalog, a visual search of
all~$\approx$~17,000 EDR spectra that were not classified as either quasars
or galaxies was completed; the spectra of~61 of these sources indicated
a possible AGN nature, and they were added to the initial quasar data base.

For the catalog we selected the subset of objects that
1)~have at least one emission line with a FWHM that exceeds
\hbox{1000 km s$^{-1}$} and 2)~have luminosities that exceed
\hbox{$M_{i^*} = -23$}.  The FWHMs of the lines were determined by performing
Gaussian fits to the line profiles.  Note that any ``narrow-lined" (Type~II)
quasars whose lines have FWHMs less than 1000~km~s$^{-1}$ will not be included
in the catalog.

The absolute magnitudes were calculated by correcting the $i^*$ 
measurement for Galactic extinction (using the maps of
Schlegel, Finkbeiner, \& Davis~1998) and assuming that the quasar
spectral energy distribution in the ultraviolet-optical
can be represented by a power law
\hbox{($f_{\nu} \propto \nu^{\alpha}$),} where $\alpha$~=~$-0.5$
(Vanden~Berk et al.~2001).
The~$i$ band was selected for the luminosity indicator rather than the
more standard definition that uses the~$B$ filter primarily because
of the ability of the SDSS to detect high-redshift quasars.  Luminosity
estimates of objects at redshifts
where the Lyman~$\alpha$ emission line is shifted redward of the observed
filter are unreliable because of the absorption produced by the
Lyman~$\alpha$ forest and Lyman-limit systems;
the Lyman~$\alpha$ line does not reach the
center of the~$i$ filter until redshifts of~$\approx$~5.  Other advantages
of the $i$ band are 1)~the flux limit of the SDSS quasar survey is
set by the~$i$ band flux and 2)~Galactic (and internal) reddening will
be less in $i$ than in~$B$ measurements.  This definition actually
matches the canonical definition of \hbox{$M_B = -23$} quite well;
for typical quasars, the rest frame \hbox{$(B - i) \approx +0.35$.}
The only significant drawback to basing quasar luminosities on~$i$ rather
than~$B$ is that at redshifts of a few tenths or less 
the luminosity calculation could be heavily influenced by
the presence of a luminous
stellar component rather than the quasar continuum (although note that
we use point spread function magnitudes in this calculation,
not Petrosian magnitudes, even for extended sources).

These criteria reduced the sample from~4564 to~3847~objects.
Of the~717 objects that were dropped from the catalog, only ten were rejected
solely because of the line width requirement.  Of these ten objects, seven 
came from the second supplemental sample, which was produced by
a visual inspection of the spectra and objects were included if the
emission line appeared to be resolved; the other three objects, all from
the original EDR quasar query, had
poor quality spectra that were erroneously assigned redshifts larger than six.
As expected, the vast majority of the rejected objects were low-luminosity
active galactic nuclei; the SDSS images of~82\% of the rejections
were morphologically classified as extended sources.

The SDSS spectrum of each of the 3847 quasars was manually inspected by
several of the authors.  The SDSS pipeline redshifts were undoubtedly
correct for over~97\% of the objects; for most of the remaining
spectra the redshift status flag indicated that the redshift was either
uncertain or, in some cases, unknown.  These spectra tended to be either
of very low signal-to-noise ratio, strong (often spectacular) broad
absorption line (BAL) quasars, or spectra containing only one, relatively
weak, emission line.

Upon review of the spectra, a consensus was reached that it was impossible
to reliably determine the redshifts of~32 objects; these were dropped
from the sample (but of course are available as part of the EDR).
The spectroscopic pipeline redshifts of~33 objects were 
significantly in error (again, most were flagged as uncertain measurements);
the revised redshifts are included in this catalog.  The revised redshift
for one of the objects caused it to fall below the quasar luminosity cutoff.
We have also
revised the redshifts of many of the high-redshift \hbox{($z > 4)$}
quasars (usually by~$\approx$~0.02), as automated redshift measurements of
these objects is difficult because the emission lines unaffected by the
Lyman~$\alpha$ forest either become inaccessible in the SDSS spectra or
are located in regions of the spectrum with low signal-to-noise ratio;
we have included the values determined
in previous publications (see Anderson et al.~2001 and references therein).

As a final note, we searched for BL~Lacs in the EDR by matching the
FIRST catalog with the
entire EDR spectroscopic data base, but after visual examination of the
spectra of the candidates we did not identify any unambiguous examples
of BL~Lacs that are not already in the catalog.

\section{Catalog Format}

The first edition of the SDSS Quasar Catalog consists of~3814 quasars.  The
catalog is written in ASCII format and is~685~kB in size.  The first~37 lines
consist of catalog documentation; this is followed by~3814 lines containing
information on the quasars.  There are~32 columns in each line; a summary
of the information is given in Table~2 (most of the catalog documentation
is a repeat of Table~2).

Notes on the catalog columns:

\noindent
1) The official names of the objects are given by the format
\hbox{SDSS Jhhmmss.ss+ddmmss.s}; only the final 18, nondegenerate
characters are given in the catalog.

\noindent
2-3) These columns contain the J2000 coordinates (Right Ascension and
Declination) in radians.  The positions for the vast majority of
the objects are accurate to~0.1$''$~rms in each coordinate; the largest
expected errors are~0.2$''$.

\noindent
4) The SDSS quasar redshifts are not determined by simply using the
the rest laboratory wavelengths of their emission lines, but according
to the {\em empirical} rest wavelengths of their emission lines based
upon the composite spectrum of Vanden Berk et al.~(2001) assuming that
[\ion{O}{3}] represents the systemic (center of mass) redshift of the
quasars.  The redshifts are so determined because it is now well-known
that the empirical centers of many quasar emission lines (the high
ionization lines in particular) are shifted with respect to the
systemic redshift of the quasars (Tytler \& Fan~1992, Vanden Berk et al. 2001,
and references therein).  The method used by the SDSS will
produce redshifts that are much closer to the systemic redshift in the
ensemble average; see Stoughton et al.~(2002) for more details regarding
how the redshift determination is implemented.
The statistical errors of the redshifts,
based on either the height and width of the
cross correlation function (c.f., Tonry \& Davis~1979) or on the
scatter of the redshifts measured from the individual emission lines,
are less than~0.01 for the non-BALs \hbox{and 0.01-0.03} for BAL quasars.

\noindent
5) The data base search technique used to find the quasar is coded in this
column. If the spectrum of the object was classified as a quasar by the
SDSS software, this column contains a ``0"; a ``1" indicates the~16 objects
identified in the extreme BAL search, a ``2" is given for the~10 quasars found
in the extensive visual search of EDR spectra that were not classified as
either quasars or galaxies by the SDSS software.

\noindent
6-15) These columns contain the
magnitudes and errors for each object in the five SDSS filters.  The
values refer to magnitudes measured by fitting to the point spread
function to the data (see Stoughton et al.~2002).
Note that the quantities are asinh magnitudes (Lupton, Gunn, \& Szalay~1999),
which are defined by

$$  m \ = \ -{2.5 \over \ln 10} \ \left[ \ {\rm asinh} \left({f/f_0 \over
2 b} \ \right) \ + \ \ln b \ \right] $$

\noindent
where $f_0$ is the flux from a zero magnitude object and the quantity
$b$ is the softening parameter.
The SDSS has set $b$, which is dimensionless, such that zero flux corresponds
to magnitudes~24.63, 25.11, 24.80, 24.36, and~22.83
in the $u$, $g$, $r$, $i$,
and~$z$ bands, respectively (Stoughton et al.~2002).
For measurements that are approximately
2.5~magnitudes brighter than the zero flux values,
the difference between asinh magnitudes and standard magnitudes (Pogson~1856)
are less than~1\%; for the vast majority of the entries in the catalog
the differences between asinh and standard magnitudes are negligible
(the primary exceptions being the $u$ magnitudes of high-redshift quasars).
The SDSS photometric system is normalized so that the $ugriz$ magnitudes are
on the AB system (Oke \& Gunn~1983).

\vbox{\noindent
16) Galactic absorption in the $u$ band based on the maps of
Schlegel, Finkbeiner, \& Davis~(1998).  For an $R_V = 3.1$ absorbing medium,
the absorptions in the SDSS bands are

$$ A_u \ = \ 5.155 \ E(B-V) $$
$$ A_g \ = \ 3.793 \ E(B-V) $$
$$ A_r \ = \ 2.751 \ E(B-V) $$
$$ A_i \ = \ 2.086 \ E(B-V) $$
$$ A_z \ = \ 1.479 \ E(B-V) $$
}

\noindent
17) If there is a source
in the FIRST catalog 
within~2.0$''$ of
the quasar position, this column contains the FIRST
peak flux density (mJy) at 20~cm.

\noindent
18) The logarithm
of the vignetting corrected count rate (photons s$^{-1}$)
in the broad energy band in the following ROSAT catalogs:
All-Sky Survey Faint Source Catalog (Voges et al.~2000); 
All-Sky Survey Bright Source Catalog (Voges et al.~1999);
and the PSPC Pointing and HRI Pointing Catalogs (private communication from
the ROSAT Result Archive collaboration).  The matching radius was set
to~60.0$''$ (Faint Source Catalog), 30$''$ (Bright Source Catalog and
PSPC Pointings) and 10$''$ (HRI Pointings); an entry of~``0"
in this column indicates no X-ray detection.

\noindent
19) The absolute magnitude in the $i$ band calculated assuming
$H_0$~=~50, $\Omega_M$~=~1, and $\Omega_{\Lambda}$~=~0, a power
law (frequency) index of~$-0.5$, and correcting the $i^*$ measurement
for Galactic extinction.

\noindent
20) If the SDSS photometric pipeline classified the image of the quasar
as a point source, the catalog entry is~0; if the quasar is extended, the
catalog entry is~1.

\noindent
21) The version of the quasar target selection algorithm used to
select the object is coded in this column
\hbox{(1 = v2.2a,}
\hbox{2 = v2.5,}
\hbox{3 = v2.7)}; see Stoughton et al.~(2002)
for details of the different techniques.

\noindent
22-27) These six columns
indicate the spectroscopic target selection status for each object.
An entry of~``1" indicates that the object satisfied the given criterion
(see Stoughton et al.~2002 for details).  Note that an object can
be targeted by more than one selection algorithm. 

\noindent
29-31) Information about the spectroscopic observation (modified Julian
date, spectroscopic plate number, and spectroscopic fiber number) used to
determine the redshift are contained
in these columns.

\noindent
32) If there is a source in the NED quasar data base within~5.0$''$ of the
quasar position, the NED object name is given in this column, unless the
NED name refers to an SDSS-discovered object.

In addition to the catalog, the SDSS spectra of all objects in the
catalog are available at a public internet site\footnote{\tt
http://archive.stsci.edu/sdss/documents/prepared\_ds.html}.

\section{Catalog Summary}

Of the 3814 objects in the catalog, 3000 were discovered by the SDSS.
The 3814 quasars span a wide range of properties: redshifts from~0.15 to~5.03,
\hbox{$ 15.16 < i^* < 20.82$} (only four objects \hbox{have $i^* > 20.5$),}
and \hbox{$ -30.1 < M_{i^*} < -23.0$.}  The catalog contains~329 matches
with ROSAT catalogs and~326 FIRST sources, as well as a number of unusual
BAL quasars.

Figure~3 displays the $i^*$-redshift relation for the quasars.  Previously
known objects are indicated with open circles.  The curved cutoff on the left
hand side of the graph is due to the minimum luminosity criterion
\hbox{($M_{i^*} < -23$).}  The ridge of points just fainter than
\hbox{$i^* = 19$} at redshifts below three is the flux limit of the
low redshift sample; low-redshift points fainter than \hbox{$i^* = 19$}
primarily represent objects selected via criteria other than the primary
multicolor sample ($e.g.,$ serendipity).  Above a redshift
of~$\approx$~3, nearly all the quasars in the catalog were discovered by
the~SDSS.

A histogram
of the catalog redshifts is shown in Figure~4.  The clear majority of
the quasars have redshifts below two (the median redshift is~1.46),
but there is a significant tail
of objects out to a redshift of five.  The dip in the curve at redshifts
between~3.3 and~3.5 is due to difficulties encountered selecting these
objects during the commissioning period; the final version of the target
selection algorithm will significantly reduce the number of missed quasars
in this redshift region.

\subsection{Analysis of Quasar Selection}

A summary of the spectroscopic selection is given in Table~3.  There are
six selection classes, which are columns~22 to~27 in the catalog.
The second column in Table~3 gives the numbers of each object that
satisfied a given selection criteria, the third column contains the number
of objects that were identified only by that selection class.
As expected, the solid majority~(81\%) were selected based on the SDSS
quasar selection criteria;
one-third of the catalog objects were
selected on that basis only.  Over~60\% of the quasars
were identified by the serendipity code, which is also primarily an ``unusual
color" algorithm.  About one-seventh of the catalog was selected by
the serendipity criteria alone; these objects tend to be low-redshift
quasars that fall below the magnitude limit of the quasar survey algorithm.

Of the~2528 quasars with \hbox{$i^* < 19.0$,}
2477 were found from the quasar multicolor
selection; if one includes multicolor and FIRST selection, then
only~20 \hbox{$i^* < 19.0$} catalog quasars are missed.  For the entire
catalog, over~99\% of the quasars are selected by either multicolor, FIRST,
or serendipity criteria.
It should be noted that the rather high fraction of faint \hbox{$z<3$}
quasars selected by serendipity is not likely to be representative of
future quasar target selection, as the number of quasar
serendipity targets in the catalog is a result of circumstances
specific to the commissioning period.  We expect that the final SDSS selection
algorithms, designed to produce complete samples of galaxies and quasars,
will have a smaller fraction of fibers assigned to serendipitous targets
than was the situation during the commissioning period.

One of the catalog entries, \hbox{SDSS J114324.97$-$003614.5,} was selected
by the ROSAT algorithm but is not listed as a ROSAT source.  This situation
arose because during the SDSS commissioning period, the ROSAT catalogs were
being upgraded; this source was included in early versions of the
ROSAT catalogs but did not meet the final ROSAT selection criteria.

\subsection{Bright Quasars}

Fourteen of the catalog quasars have \hbox{$i^* < 16.5$}, but only one,
\hbox{SDSS J014942.50+001501.7,} a \hbox{$i^* = 16.24$} quasar
at a redshift of~0.55, was not previously known (see spectrum in Figure~2).
The catalog contains~51
quasars brighter than \hbox{$i^* = 17.0$}; 16 are SDSS discoveries.
Four of the sixteen SDSS discoveries are FIRST sources, and seven have been
detected with ROSAT.

\subsection{Luminous Quasars}

Of the nine catalog quasars with \hbox{$M_{i^*} < -28.8$} (3C~273 has
\hbox{$M_{i} = -27.3$} in our adopted cosmology), three were not previously
known.
\hbox{SDSS J173352.23+540030.5}, at \hbox{$z = 3.43$} and
\hbox{$M_{i^*} = -29.4$,} is the second
most luminous quasar (after HS~1700+6416 = \hbox{SDSS J170100.62+641209.0)}
in the catalog;
its spectrum is displayed in Figure~2.  This object is a~7.72~mJy FIRST
source.
The other new highly luminous quasars are \hbox{SDSS J012412.47$-$010049.8}
($z$~=~2.82, $M_{i^*}$~=~$-29.4$), and
\hbox{SDSS J152119.68$-$004818.8} \hbox{($z = 2.93$,}
\hbox{$M_{i^*} = -28.9$).}

\subsection{Broad Absorption Line Quasars}

The catalog contains a number of BAL quasars, and it is clear that the SDSS
is quite effective at finding extreme examples of this class (see
Hall et al.~2002 for a discussion).  One of the more spectacular examples
is \hbox{SDSS J172341.09+555340.5}, a $z$~=~2.113 low-ionization BAL quasar
whose troughs have a ``scalloped" appearance (see Figure~2).

\subsection{Low-Redshift Quasars}

The quasar with the lowest redshift in the catalog,
\hbox{SDSS J232259.99$-$005359.3} at \hbox{$z = 0.15$}, was not previously
known (spectrum is shown in Figure~2).  Of course, there are many
lower-redshift AGN in the EDR that do not satisfy our luminosity
criterion. 
Of the~49 quasars with redshifts below~0.30, 33 are reported here
for the first time, as well as eight new \hbox{$z < 0.20$} quasars.
Included in the new discoveries are five FIRST sources and fifteen
ROSAT detections.
One of the low-redshift objects,
\hbox{SDSS J011254.91+000313.0} at \hbox{$z = 0.24$}, is PB~06317 in the NED
data base with the erroneous redshift of \hbox{$z = 1.23$.}

\subsection{High-Redshift Quasars}

Two of the fifty $z > 3.95$ quasars in the catalog have not been
previously published: 
\hbox{SDSS J103432.71$-$002702.5}, at \hbox{$z = 4.38$}, and
\hbox{SDSS J122657.97+000938.4}, at \hbox{$z = 4.14$}.  The spectra
of both objects are displayed in Figure~2.  Observations of most of the
\hbox{$z > 4$} quasars in the catalog were described by Anderson et al.~(2001).
Although only a small fraction of the objects in the catalog are at
high-redshift, the SDSS is clearly very effective at discovering
such quasars; 96 of the~101 catalog entries with \hbox{$z > 3.5$}
are SDSS discoveries.

\subsection{Close Pairs}

There are 14 pairs of quasars in the catalog with angular separation less than
$70''$.  One pair, \hbox{SDSS J025959.69+004813.5} and
\hbox{SDSS J030000.56+004828.0}, both new discoveries, have redshifts
($z=0.893$) identical
within the errors, and a separation of only~$19.5''$,
corresponding to a physical distance of 164~kpc;
this is almost certainly a binary quasar. 

\subsection{Redshift Disagreements with Previous Measurements}

The redshifts of twenty quasars in this catalog disagree by more than~0.07
from the values given in the NED database; the information for each
of these objects is given in Table~4.  All of the redshift discrepancies
are larger than~0.4, and in ten of the cases the SDSS redshift is smaller than
the NED.  Three of the quasars in Table~4 have blank fields in the
NED redshift column (identified by \hbox{``..."} in the third column of
Table~4), and seven of the Table~4 objects are
from the UM survey.  The incorrect published redshifts of UM~203 and UM~183
were noted by Richards et al.~(2001a).

We have reexamined the spectra of the~17 objects with actual redshift
disagreements, and for all but a few the SDSS value is undoubtedly the
correct one; in the cases that are not certain, we believe the SDSS
redshifts are the more likely ones. Examples where the SDSS redshift could
be incorrect include
the two BAL quasars \hbox{LBQS 0021$-$0100} and \hbox{[HB89]2345+002,}
and the 2dF survey object \hbox{2QZ J130916.6$-$001550,}
where Croom et al.~(2001) identify the emission line at~6900~\AA\ as
Mg~II, whereas we interpret this line as H$\beta$ and the feature
near~4000~\AA\ as Mg~II emission.  Note that quality flag
for \hbox{2QZ J130916.6$-$001550} given by Croon et al.~(2001) indicates an
uncertain redshift.

\subsection{Morphology}

The SDSS photometric pipeline classifies the images of~122 quasars as
resolved.  As one would expect, the vast majority (87\%)
of the extended objects
have redshifts below one, but there are a number of resolved quasars at higher
redshifts (the highest redshift of an extended object is~3.61).
The majority of the large redshift ``resolved" quasars are probably measurement
errors, but this sample probably contains a few chance superpositions
of quasars and foreground objects or possibly some
small angle separation gravitational lenses.  A detailed study of
one of the extended
\hbox{$z > 1$} quasars in the catalog, \hbox{SDSS J122608.02$-$000602.2}
at a redshift of~1.12,
shows that the source consists of a close pair of quasars and is likely to be a
gravitational lens system (Inada et al.~2002).

\section{Conclusion}

The current catalog contains slightly less than 5\% of the planned survey
area, although again we point out that the observations in the EDR were
taken during the commissioning period and the final algorithm will provide
more systematic study of the survey region.
We plan to produce regular updates to
the quasar catalog, on approximately
one year intervals, throughout the duration of the survey.  The next release
is scheduled for late~2002; we expect it to contain more than 20,000
quasars, most identified with the final SDSS quasar selection algorithm.

\acknowledgments

We would like to thank Niel Brandt for his assistance with preparing
the X-ray information for the catalog, and Megan Donahue for preparing the
public web site.
This work was supported in part by National Science Foundation grants
AST99-00703~(DPS and~GTR), PHY00-70928~(XF), and AST00-71091~(MAS).
XF and MAS acknowledge
additional support from the Princeton University
Research Board and a Porter O.~Jacobus Fellowship, and XF acknowledges
support from a Frank and Peggy Taplin Fellowship.

The Sloan Digital Sky Survey
\footnote{The SDSS Web site \hbox{is {\tt http://www.sdss.org/}.}}
(SDSS) is a joint project of The University
of Chicago, Fermilab, the Institute for Advanced Study, the Japan
Participation Group, The Johns Hopkins University, the
Max-Planck-Institute for Astronomy (MPIA), the Max-Planck-Institute for
Astrophysics
(MPA), New Mexico State University, Princeton University, the United
States Naval Observatory, and the University of Washington. Apache
Point Observatory, site of the SDSS telescopes, is operated by the
Astrophysical Research Consortium (ARC). 
Funding for the project has been provided by the Alfred~P.~Sloan
Foundation, the SDSS member institutions, the National Aeronautics and
Space
Administration, the National Science Foundation, the U.S.~Department of
Energy, the Japanese Monbukagakusho, and the Max Planck Society.

This research has made use of the NASA/IPAC Extragalactic Database (NED)
which is operated by the Jet Propulsion Laboratory, California Institute
of Technology, under contract with the National Aeronautics and Space
Administration.

\newpage
\centerline{\bf Figure Captions}

\figcaption{
Celestial locations (equatorial, J2000)
of the~92 spectroscopic plates used for the catalog.
Each plate has a diameter of~3$^{\circ}$.  The total area covered by the
plates is 494~deg$^2$.
\label{fig1}}

\figcaption{
An example of data produced by the SDSS spectrographs.  The spectral
resolution of the data is~$\approx$~1800; a dichroic splits the beam
at~6150~\AA .  The data have been rebinned \hbox{to 4 \AA\ pixel$^{-1}$}
for display purposes.  All six of the quasars were discovered by the~SDSS.
Notes on spectra: 1)~\hbox{SDSS J232259.99$-$005359.3} is the
lowest redshift object in the catalog; 2)~\hbox{SDSS J014942.50+001501.7}
is the brightest \hbox{($i^* = 16.24$)} new object in the catalog;
3)~\hbox{SDSS J173352.23+540030.5}, with \hbox{$M_{i^*} = -29.4$}, is the
second most luminous quasar in the catalog and the most luminous new object;
4)~\hbox{SDSS J172341.09+555340.5} is an example of an extreme~BAL;
5,6)~\hbox{SDSS J103432.71$-$002702.5} and
\hbox{SDSS J122657.97+000938.4} are the two new \hbox{$z > 4$} quasars in the
catalog.
\label{fig2}}

\figcaption{
A plot of the observed~$i^*$ magnitude as a function of redshift for the
3814 objects in the catalog.  Open circles indicate quasars not discovered by
the SDSS.
\label{fig3}}

\figcaption{
The redshift histogram of the catalog quasars.  The smallest redshift is~0.15
and the largest redshift is~5.03; the median redshift of the catalog is~1.46.
\label{fig4}}

\clearpage

%

\begin{figure}
\plotfiddle{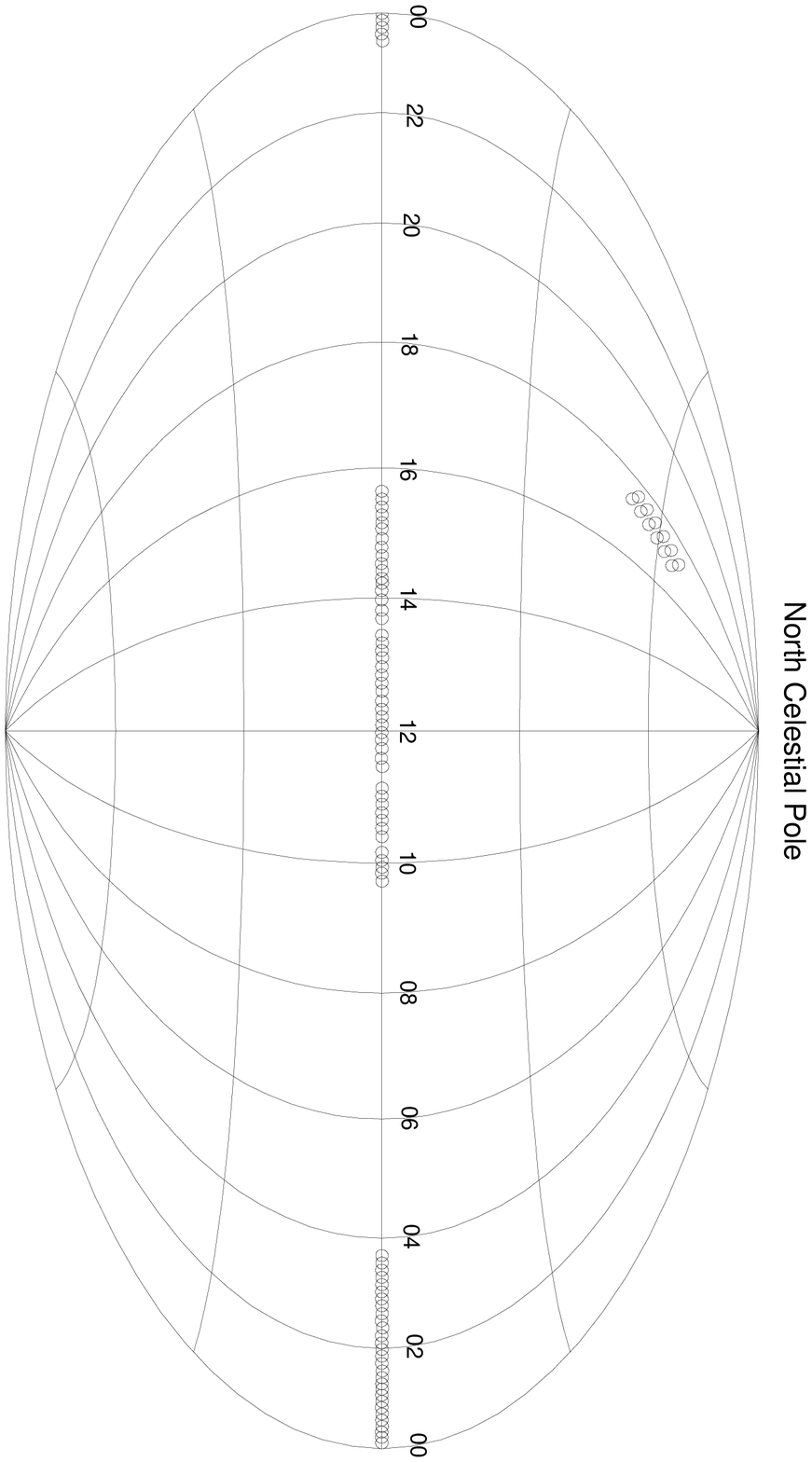}{8.0in}{180.0}{85.0}{85.0}{250.0}{600.0}
\label{Figure 2 }
\end{figure}

\begin{figure}
\plotfiddle{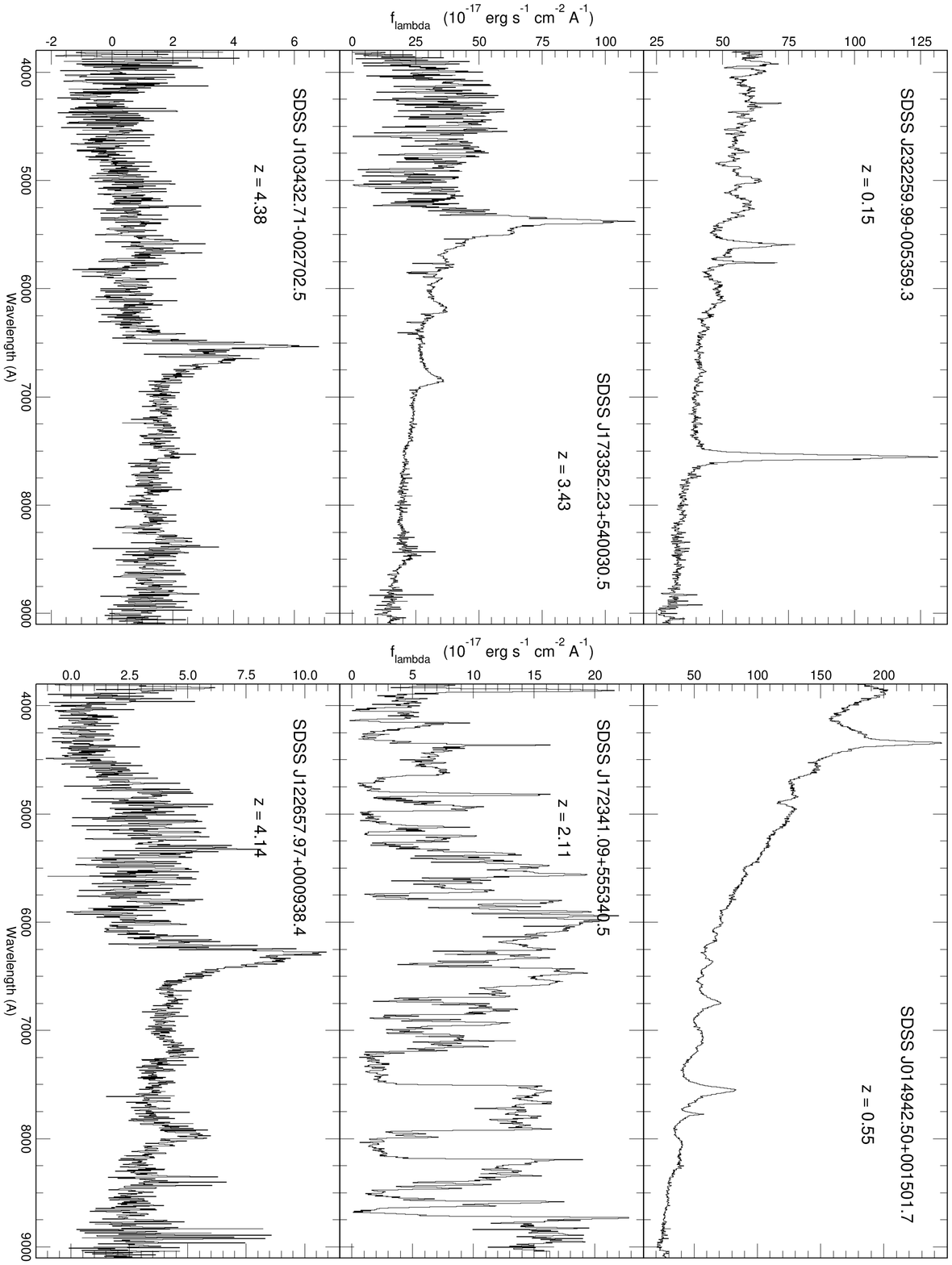}{8.0in}{180.0}{80.0}{80.0}{250.0}{600.0}
\label{Figure 3 }
\end{figure}

\clearpage

\begin{figure}
\plotfiddle{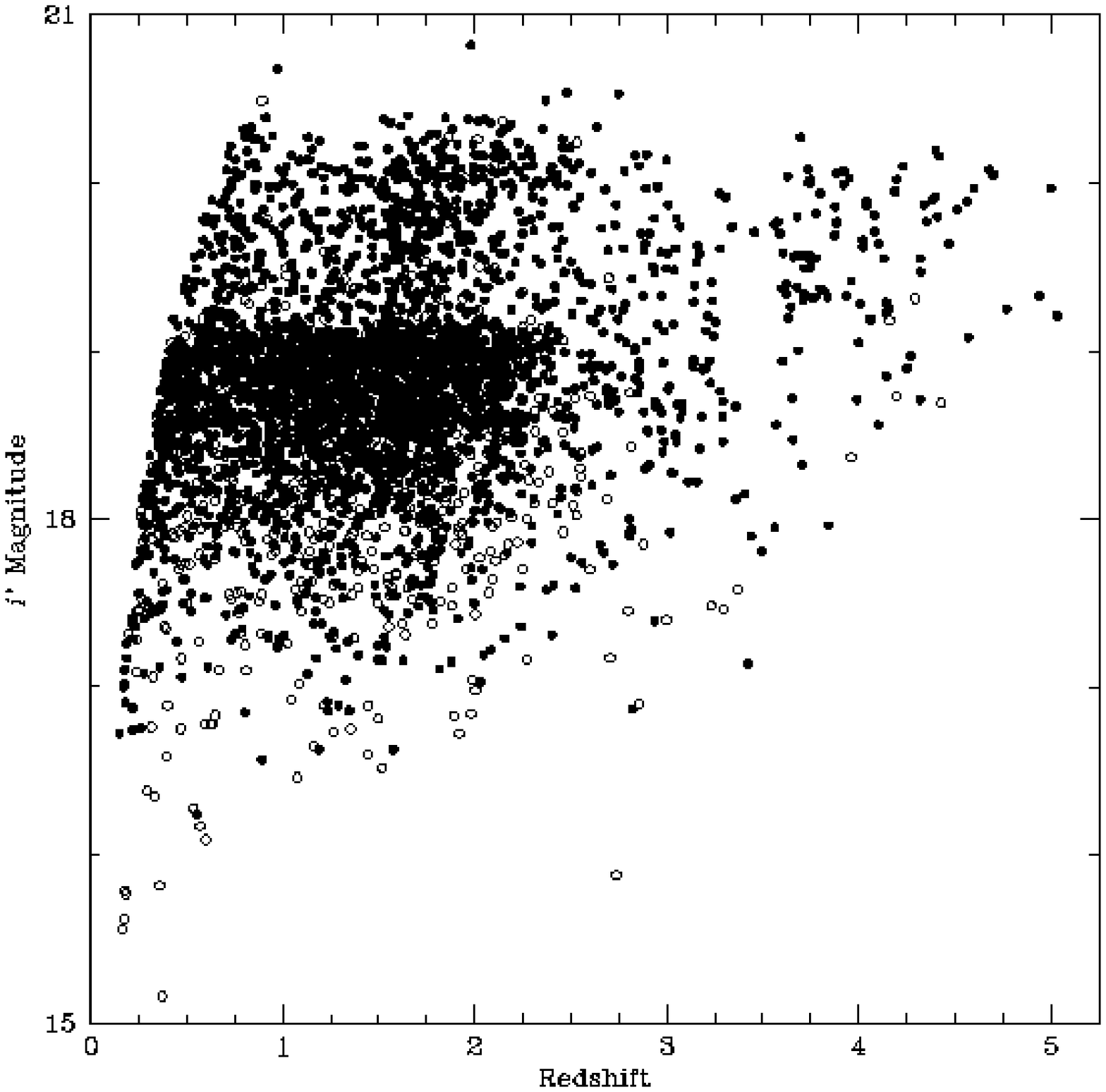}{8.0in}{0.0}{90.0}{90.0}{-270.0}{-110.0}
\label{Figure 4 }
\end{figure}

\clearpage

\begin{figure}
\plotfiddle{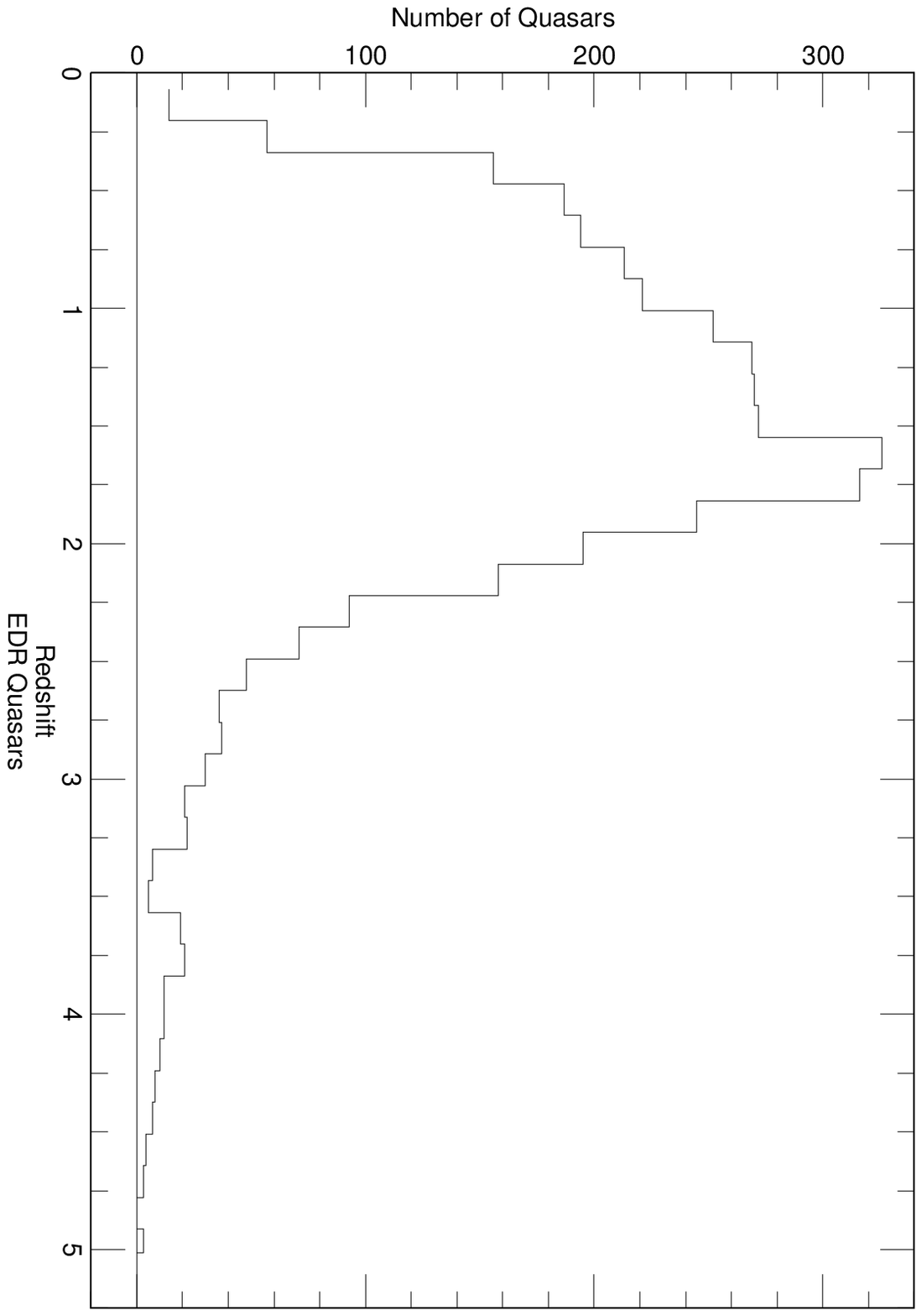}{8.0in}{180.0}{110.0}{110.0}{210.0}{825.0}
\label{Figure 5 }
\end{figure}

\clearpage


\begin{footnotesize}
\vbox{
\halign{\hskip 12pt
\hfil # \hfil \tabskip=1em plus1em minus1em&
\hfil # \hfil &
\hfil # \hfil &
\hfil # &
# &
\hfil # \hfil &
\hfil # \hfil &
\hfil # \hfil &
\hfil # \cr
\multispan8{\hfil TABLE 1 \hfil} \cr
\noalign{\smallskip}
\multispan8{\hfil SDSS Spectroscopic Plate Information
 \hfil}\cr
\noalign{\smallskip\hrule\smallskip\hrule\smallskip}
\hfil Plate \hfil & \hfil $\alpha_{2000}$ \hfil &
 \hfil $\delta_{2000}$ \hfil & \hfil N \hfil & \ \ \ \ \ &
\hfil Plate \hfil & \hfil $\alpha_{2000}$ \hfil &
 \hfil $\delta_{2000}$ \hfil & \hfil N \hfil \cr
\noalign{\smallskip\hrule\medskip}
  266 & 09 43 34.3 & $+$00 03 41 &  24 &&
  349 & 16 56 34.6 & $+$63 39 11 &  26 \cr
  267 & 09 50 56.0 & $-$00 01 50 &  18 &&
  350 & 17 13 59.4 & $+$65 08 01 &  40 \cr
  268 & 09 56 14.8 & $+$00 04 47 &  12 &&
  351 & 17 03 48.5 & $+$61 37 10 &  42 \cr
  269 & 10 02 31.1 & $+$00 00 00 &  23 &&
  352 & 17 20 00.3 & $+$63 02 14 &  35 \cr
  270 & 10 09 50.6 & $+$00 00 22 &  17 &&
  353 & 17 10 08.7 & $+$59 33 53 &  38 \cr
  272 & 10 24 04.5 & $+$00 01 06 &  41 &&
  354 & 17 25 12.6 & $+$60 55 38 &  75 \cr
  273 & 10 31 36.5 & $+$00 00 33 &  45 &&
  355 & 17 15 44.9 & $+$57 29 35 &  33 \cr
  274 & 10 38 59.7 & $+$00 01 06 &  30 &&
  358 & 17 33 47.8 & $+$56 40 34 &  71 \cr
  275 & 10 45 51.9 & $-$00 00 22 &  29 &&
  359 & 17 25 14.8 & $+$53 18 40 &  71 \cr
  276 & 10 53 31.3 & $+$00 02 34 &  56 &&
  360 & 17 37 23.6 & $+$54 32 20 &  69 \cr
  277 & 11 00 55.2 & $+$00 00 22 &  38 &&
  366 & 17 29 45.9 & $+$58 48 22 &  73 \cr
  278 & 11 08 06.6 & $+$00 01 39 &  42 &&
  367 & 17 20 44.9 & $+$55 24 28 &  40 \cr
  281 & 11 27 30.5 & $+$00 06 37 &  34 &&
  383 & 23 24 48.5 & $+$00 06 51 &  50 \cr
  282 & 11 35 04.8 & $-$00 07 10 &  60 &&
  384 & 23 33 10.8 & $-$00 02 23 &  40 \cr
  283 & 11 43 54.0 & $-$00 00 11 &  54 &&
  385 & 23 41 33.5 & $+$00 02 40 &  55 \cr
  284 & 11 51 55.4 & $-$00 03 52 &  42 &&
  386 & 23 50 31.7 & $+$00 03 15 &  66 \cr
  285 & 11 58 14.6 & $-$00 00 55 &  29 &&
  387 & 23 59 17.8 & $+$00 01 27 &  41 \cr
  286 & 12 05 13.5 & $+$00 00 33 &  38 &&
  388 & 00 07 17.9 & $-$00 00 18 &  32 \cr
  287 & 12 12 37.4 & $+$00 01 06 &  36 &&
  389 & 00 14 14.4 & $-$00 01 29 &  41 \cr
  288 & 12 19 23.0 & $-$00 01 06 &  56 &&
  390 & 00 20 53.5 & $-$00 02 00 &  35 \cr
  289 & 12 26 49.9 & $-$00 01 06 &  53 &&
  391 & 00 27 59.1 & $+$00 02 43 &  20 \cr
  290 & 12 35 43.5 & $+$00 01 50 &  57 &&
  392 & 00 35 35.0 & $-$00 00 08 &  39 \cr
  291 & 12 42 54.2 & $-$00 02 12 &  26 &&
  393 & 00 43 16.5 & $+$00 02 13 &  45 \cr
  292 & 12 50 19.5 & $-$00 01 28 &  50 &&
  394 & 00 50 54.6 & $-$00 01 59 &  37 \cr
  293 & 12 57 49.3 & $+$00 00 11 &  66 &&
  395 & 00 58 26.2 & $+$00 00 57 &  30 \cr
  294 & 13 06 01.8 & $+$00 05 20 &  51 &&
  396 & 01 05 52.9 & $-$00 00 14 &  34 \cr
  295 & 13 12 23.9 & $-$00 01 06 &  30 &&
  397 & 01 13 04.8 & $+$00 01 06 &  35 \cr
  296 & 13 19 28.7 & $+$00 04 03 &  32 &&
  398 & 01 19 20.5 & $+$00 00 08 &  32 \cr
  297 & 13 26 15.8 & $-$00 00 22 &  26 &&
  399 & 01 26 40.8 & $+$00 02 15 &  31 \cr
  299 & 13 41 25.7 & $+$00 00 22 &  33 &&
  400 & 01 34 32.8 & $+$00 05 30 &  43 \cr
  300 & 13 48 54.0 & $-$00 00 44 &  45 &&
  401 & 01 43 17.9 & $-$00 01 15 &  42 \cr
  301 & 13 58 28.8 & $-$00 03 19 &  54 &&
  402 & 01 51 37.5 & $+$00 00 07 &  35 \cr
  302 & 14 06 53.0 & $-$00 02 57 &  42 &&
  403 & 01 59 16.8 & $+$00 00 37 &  38 \cr
  303 & 14 13 40.1 & $+$00 04 03 &  22 &&
  404 & 02 06 35.0 & $-$00 02 07 &  22 \cr
  304 & 14 18 03.1 & $+$00 00 00 &  29 &&
  405 & 02 14 43.9 & $-$00 01 58 &  47 \cr
  305 & 14 25 13.7 & $+$00 02 01 &  28 &&
  406 & 02 23 31.9 & $+$00 07 30 &  67 \cr
  306 & 14 31 31.4 & $-$00 04 58 &  34 &&
  407 & 02 31 23.0 & $-$00 03 34 &  52 \cr
  307 & 14 38 53.9 & $+$00 03 19 &  28 &&
  408 & 02 39 17.4 & $+$00 02 13 & 123 \cr
  308 & 14 46 34.0 & $-$00 03 41 &  39 &&
  409 & 02 47 59.6 & $+$00 00 14 &  46 \cr
  309 & 14 54 31.7 & $+$00 01 28 &  43 &&
  410 & 02 55 26.8 & $-$00 01 01 &  91 \cr
  310 & 15 03 26.8 & $+$00 00 22 &  39 &&
  411 & 03 03 08.9 & $-$00 01 01 &  34 \cr
  311 & 15 09 07.7 & $+$00 01 28 &  12 &&
  412 & 03 10 56.2 & $+$00 00 13 &  35 \cr
  312 & 15 16 18.4 & $+$00 01 28 &  26 &&
  413 & 03 18 44.8 & $+$00 00 55 &  50 \cr
  313 & 15 23 37.1 & $-$00 01 17 &  24 &&
  414 & 03 26 31.5 & $-$00 00 43 &  47 \cr
  314 & 15 30 58.8 & $+$00 00 44 &  10 &&
  415 & 03 34 05.5 & $+$00 01 50 &  46 \cr
  315 & 15 38 17.6 & $-$00 00 11 &  27 &&
  416 & 03 41 58.0 & $+$00 00 50 &  70 \cr
\noalign{\smallskip\hrule}}
}
\end{footnotesize}

\clearpage

\halign{\hskip 12pt
\hfil # \tabskip=1em plus1em minus1em&
\hfil # \hfil &
# \hfil \cr
\multispan3{\hfil TABLE 2 \hfil} \cr
\noalign{\medskip}
\multispan3{\hfil Quasar Catalog Format \hfil} \cr
\noalign{\bigskip\hrule\smallskip\hrule\medskip}
\hfil Column \hfil & \hfil Format \hfil & \hfil Description \hfil \cr
\noalign{\medskip\hrule\bigskip}
   1  &   A18 &   
SDSS Object Name   \ \ \ \ hhmmss.ss+ddmmss.s  \ \ \ (J2000) \cr
   2  &  I10  &   10000000 $\times$ R.A. in radians (J2000) \cr
   3  &  I10  &   10000000 $\times$ Declination in radians (J2000) \cr
   4  &   I5  &   1000 $\times$ redshift \cr
   5  &   I2  &   0 \ =  \ EDR Quasar \ \ \ 1 \ = \ Extreme BAL Search
\ \ \ 2 \ = \ Visual Search \cr
   6  &   I5  &    100 $\times$ PSF $u^*$ magnitude \cr
   7  &   I4  &    100 $\times$ error in PSF $u^*$ magnitude \cr
   8  &   I5  &    100 $\times$ PSF $g^*$ magnitude \cr
   9  &   I4  &    100 $\times$ error in PSF $g^*$ magnitude \cr
  10  &   I5  &    100 $\times$ PSF $r^*$ magnitude \cr
  11  &   I4  &    100 $\times$ error in PSF $r^*$ magnitude \cr
  12  &   I5  &    100 $\times$ PSF $i^*$ magnitude \cr
  13  &   I4  &    100 $\times$ error in PSF $i^*$ magnitude \cr
  14  &   I5  &    100 $\times$ PSF $z^*$ magnitude \cr
  15  &   I4  &    100 $\times$ error in PSF $z^*$ magnitude \cr
  16  &   I5  &    100 $\times$ Galactic absorption in $u$ band \cr
  17  &   I7  &    100 $\times$ FIRST Peak flux density at 20 cm (mJy) \cr
  18  &   I5  &   $-$1000 $\times$ log ROSAT full band count rate \cr
  19  &   I5  &   $-$100 $\times$ $M_{i^*}$ \cr
  20  &   I3  &   Morphology flag \ \ \ 0 = point source \ \ \ 1 = extended \cr
  21  &   I3  &   Quasar Target Selection Algorithm: \ 1 \ = \ v2.2a
\ \ \ 2 \ = \ v2.5 \ \ \ 3 = v2.7 \cr
  22  &   I3  &   Spectroscopic Target flag: Multicolor Quasar (0 or 1) \cr
  23  &   I3  &   Spectroscopic Target flag: FIRST (0 or 1) \cr
  24  &   I3  &   Spectroscopic Target flag: ROSAT (0 or 1) \cr
  25  &   I3  &   Spectroscopic Target flag: Serendipity (0 or 1) \cr
  26  &   I3  &   Spectroscopic Target flag: Star (0 or 1) \cr
  27  &   I3  &   Spectroscopic Target flag: Galaxy (0 or 1) \cr
  28  &   I6  &   SDSS Imaging Run Number for photometric measurements \cr
  29  &   I6  &   Modified Julian Date of spectroscopic observation \cr
  30  &   I5  &   Spectroscopic Plate Number \cr
  31  &   I4  &   Spectroscopic Fiber Number \cr
  32  &   A20 &   Object Name for previously known quasars \cr
\noalign{\medskip\hrule}}

\clearpage

\halign{\hskip 12pt
# \hfil \tabskip=1em plus1em minus1em&
\hfil # &
\hfil # \cr
\multispan3{\hfil TABLE 3 \hfil} \cr
\noalign{\medskip}
\multispan3{\hfil Spectroscopic Target Selection \hfil} \cr
\noalign{\bigskip\hrule\smallskip\hrule\medskip}
&&\hfil Sole \hfil \cr
\hfil Class \hfil & \hfil Selected \hfil &  Selection  \cr
\noalign{\medskip\hrule\bigskip}
Quasar & 3107 & 1277 \cr
FIRST  & 223 & 18 \cr
ROSAT  & 158 & 8 \cr
Serendipity & 2334 & 549 \cr
Star & 568 & 8 \cr
Galaxy & 15 & 1 \cr
\noalign{\medskip\hrule}}

\clearpage

\halign{\hskip 12pt
# \hfil \tabskip=1em plus1em minus1em&
\hfil # \hfil &
\hfil $#$ \hfil &
# \hfil \cr
\multispan4{\hfil TABLE 4 \hfil} \cr
\noalign{\medskip}
\multispan4{\hfil Discrepant Redshifts \hfil} \cr
\noalign{\bigskip\hrule\smallskip\hrule\medskip}
\hfil Quasar (SDSS) \hfil & \hfil $z_{\rm SDSS}$ \hfil &
\hfil  z_{\rm NED} - z_{\rm SDSS} & NED Object Name \cr
\noalign{\medskip\hrule\bigskip}
J000807.53$+$001619.0 & 1.48 & -1.06 & UM 203   \cr           
J002411.66$-$004348.1 & 1.79 & -1.02 & LBQS 0021$-$0100      \cr
J003005.05$+$002848.1 & 1.41 & +0.56 & UM 248              \cr
J004319.74$+$005115.4 & 0.31 & +1.69 & UM 269              \cr
J011254.91$+$000313.0 & 0.24 & +0.99 & PB 06317            \cr
\noalign{\smallskip}
J013352.65$+$011345.1 & 0.31 & +1.06 & UM 338              \cr
J021558.50$-$005120.4 & 1.66 & +0.71 & UM 414              \cr
J021612.20$-$010518.9 & 1.49 & +0.66 & UM 416              \cr
J024840.98$-$001229.0 & 1.20 & +0.48 & US 3186             \cr
J105907.67$+$010303.4 & 1.34 & $...$ & [CCH91] 1056.6+0119 \cr
\noalign{\smallskip}
J110226.29$+$003553.0 & 0.93 & $...$ & [CCH91] 1059.9+0052 \cr
J120548.48$+$005343.8 & 0.93 & -0.83 & [HB89] 1203+011     \cr
J130916.67$-$001550.2 & 0.42 & +1.13 & 2QZ J130916.6$-$001550 \cr
J131028.50$+$004408.9 & 1.60 & $...$ & PKS B1307+010       \cr
J234340.34$+$011254.4 & 1.95 & -0.55 & [HB89] 2341+009     \cr
\noalign{\smallskip}
J234506.32$+$010135.5 & 1.79 & +0.91 & [HB89] 2342+007     \cr
J234724.71$+$005246.8 & 1.33 & -0.93 & [HB89] 2344+006     \cr
J234812.39$+$002939.5 & 1.95 & +1.11 & [HB89] 2345+002 \cr
J235008.88$-$002912.6 & 1.14 & -0.69 & UM 183              \cr
J235400.41$+$010123.4 & 1.59 & -1.16 & ZC 2351+007A        \cr
\noalign{\medskip\hrule}}

\clearpage

\clearpage


\begin{thebibliography}{}

\bibitem[Anderson et al. 2001]{sa2001}
     Anderson, S.F., Fan, X., Richards, G.T., Schneider, D.P.,
     Strauss, M.A., Vanden~Berk, D.E., et al. 2001, AJ, 122, 503
\bibitem[Becker et al 1995]{bwh95}
     Becker, R.H., White, R.L., \& Helfand, D.J. 1995, ApJ, 450, 559
\bibitem[Blanton et al. 2001]{mrb01}
      Blanton, M.R., Lupton, R.H., Maley, F.M., Young, N., Zehavi, I.,
      \& Loveday, J. 2001, AJ, submitted
\bibitem[Budavari et al. 2001]{tb01}
      Budavari, T., Csabai, I., Szalay, A.S., Connolly, A.J., et al. 2001,
      AJ, 122, in press
\bibitem[Castander et al. 2001]{fjc01}
     Castander, F.J., Nichol, R.C., Merelli, A., Burles, S., Pope, A., et al.
     2001, AJ, 121, 2331
\bibitem[Croom et al. 2001]{csb01}
     Croom, S.M., Smith, R.J., Boyle, B.J., Shanks, T., Loaring, N.S.,
     Miller, L., \& Lewis, I.J., MNRAS, in press
\bibitem[Fan et al. 2001]{fan01}
     Fan, X., Narayanan, V.K., Lupton, R.H., Strauss, M.A., et al. 2001,
     AJ, 122, in press
\bibitem[Fukugita et al. 1996]{fig96}
     Fukugita, M., Ichikawa, T., Gunn, J.E., Doi, M., Shimasaku, K.,
     \& Schneider, D.P. 1996, AJ, 111, 1748
\bibitem[Gunn et al. 1998]{gcam98}
     Gunn, J.E., Carr, M.A., Rockosi, C.M., Sekiguchi, M., et al. 1998,
     AJ, 116, 3040
\bibitem[Hall et al. 2002]{pbh02}
      Hall, P.B., Anderson, S.F., et al. 2002, in preparation
\bibitem[Hogg et al. 2001]{hsf01}
     Hogg, D.W., Schlegel, D.J., Finkbeiner, D.P., \& Gunn, J.E. 2001,
     AJ, 122, in press
\bibitem[Inada et al. 2002]{ina02}
     Inada, N., Sekiguchi, M., Anderson, S.F., Burles, S., et al. 2002, in
     preparation
\bibitem[Lupton et al. 2001]{lgi01}
     Lupton, R.H., Gunn, J.E., Ivezi\'c, \v{Z}., Knapp, G.R., Kent, S., \&
     Yasuda, N. 2001, Proceedings of ADASS X, in press
\bibitem[Lupton, Gunn, \& Szalay 1999]{lgs99}
     Lupton, R.H., Gunn, J.E., \& Szalay, A. 1999, AJ, 118, 1406
\bibitem[Oke \& Gunn 1983]{og83}
     Oke, J.B., and Gunn, J.E., 1983, ApJ., 266, 713
\bibitem[Pogson 1856]{nrp56}
     Pogson, N.R., 1856, MNRAS, 17, 12
\bibitem[Richards et al 2002]{gtr02}
      Richards, G.T., Fan, X., Newberg, H.J., Strauss, M.A., et al. 2002,
      in preparation
\bibitem[Richards et al 2001a]{gtr01a}
      Richards, G.T., Fan, X., Schneider, D.P., Vanden~Berk, D.E.,
      Strauss, M.A., et al. 2001a, AJ, 121, 2308
\bibitem[Richards et al 2001b]{gtr01b}
      Richards, G.T., Weinstein, M.A., Schneider, D.P., Fan, X.,
      Strauss, M.A., Vanden~Berk, D.E., et al. 2001b, AJ, 122, in press
\bibitem[Schlegel et al. 1998]{djs98}
      Schlegel, D.J., Finkbeiner, D.P., \& Davis, M. 1998, ApJ, 500, 525
\bibitem[Schmidt, M. 1963]{ms63}
      Schmidt, M. 1963, Nature, 197, 1040
\bibitem[Sharp et al. 2001]{rgs01}
      Sharp, R.G., McMahon, R.G., Irwin, J.J., \& Hodgkin, S.T., 2001,
      MNRAS, in press
\bibitem[Stoughton et al. 2002]{cs01}
       Stoughton, C., Lupton, R.H., et al. 2002, AJ, submitted
\bibitem[Tonry \& Davis 1979]{td79}
       Tonry, J., and Davis, M. 1979, ApJ, 84, 1115
\bibitem[Tytler \& Fan 1992]{tf92}
       Tytler, D., \& Fan, X. 1992, ApJS, 79, 1
\bibitem[Vanden Berk et al. 2001]{dvb01}
     Vanden Berk, D.E., Richards, G.T., Bauer, A., Strauss, M.A., et al. 2001,
     AJ, 122, 549
\bibitem[Veron-Cetty \& Veron 2001]{vv01}
     V\'eron-Cetty, M.P., \& V\'eron, P. 2001, A \& A, 374, 92
\bibitem[Voges et al. 1999]{voges99}
     Voges, W., et al. 1999, A \& A, 349, 389
\bibitem[Voges et al. 2000]{voges00}
     Voges, W., et al. 2000, IAUC, 7432
\bibitem[York et al.~2000]{york00}
     York, D.G., Adelman, J., Anderson, J.E., Anderson, S.F., et al. 2000,
     AJ, 120, 1579
\bibitem[Zheng et al. 2000]{wz00}
     Zheng, W., Tsvetanov, Z.I., Schneider, D.P., Fan, X., et al. 2000,
     AJ, 120, 1607
     
\end{thebibliography}
\end{document}